\vspace{-20pt}
\documentclass[journal]{IEEEtran}
\usepackage{balance}
\vspace{-20pt}
\ifCLASSINFOpdf
\else
   \usepackage[dvips]{graphicx}
\fi
\usepackage{subfigure}
\usepackage{cite}
\usepackage{mathrsfs}
\usepackage{amsmath,amssymb,amsfonts}
\usepackage{url}
\usepackage{algorithmic}
\usepackage{lipsum}
\usepackage{multicol}
\usepackage{graphicx}
\usepackage{epstopdf}
\usepackage{textcomp}
\usepackage{caption}
\usepackage{graphicx}
\usepackage{algorithmic}
\usepackage{algorithm}
\usepackage{flushend}
\usepackage{bm}
\usepackage{array}
\usepackage{setspace}
\usepackage{makecell,multirow,diagbox}
\usepackage{textcomp,booktabs}
\usepackage[usenames,dvipsnames]{color}
\usepackage{colortbl}
\definecolor{mygray}{gray}{.9}
\definecolor{mypink}{rgb}{.99,.91,.95}
\definecolor{mycyan}{cmyk}{.3,0,0,0}

\ifCLASSOPTIONcompsoc
\usepackage[caption=true, font=footnotesize, labelfont=sf, textfont=sf]{subfig}
\else
\usepackage[caption=true, font=footnotesize]{subfig}
\begin{document}

\title{\vspace{-0.5em}\LARGE Trajectory and Transmit Power Optimization for IRS-Assisted UAV Communication under Malicious Jamming }
	
\author{Zhi Ji,
	Wendong Yang,
	Xinrong Guan,	
	Xiao Zhao,
	Guoxin Li,
	and Qingqing Wu\vspace{-2em}
\thanks{
	 Zhi Ji, Wendong Yang, Xinrong Guan, Xiao Zhao, and Guoxin Li are with the College of Communications Engineering, Army Engineering University of PLA, Nanjing, 210007, China (e-mail: jz20211009@163.com; ywd1110@163.com; {guanxr@aliyun.com}; zhaoxiao1982lgd@163.com;  {gxl$\_$li@sina.com}). Q. Wu is with the State Key Laboratory of Internet of Things for Smart City, University of Macau, Macau, 999078, and also with the National Mobile Communications Research Laboratory, Southeast University, Nanjing 210096, China (email: qingqingwu@um.edu.mo).   
}
}


\maketitle
\begin{abstract}
In this letter, we investigate an unmanned aerial vehicle (UAV) communication system, where an intelligent reflecting surface (IRS) is deployed to assist in the transmission from a ground node (GN) to the UAV in the presence of a jammer. We aim to maximize the average rate of the UAV commnunication by jointly optimizing the GN's transmit power, the IRS's passive beamforming and the UAV's trajectory. However, the formulated problem is difficult to solve due to the non-convex objective function and the coupled optimization variables. Thus, to tackle it, we propose an alternating optimization (AO) based algorithm by exploiting the successive convex approximation (SCA) and semidefinite relaxation (SDR) techniques. Simulation results show that the proposed algorithm can significantly improve the average rate compared with the benchmark algorithms. Moreover, it also shows that when the jamming power is large and the number of IRS elements is relatively small, deploying the IRS near the jammer outperforms deploying it near the GN, and vice versa.
\end{abstract}


\IEEEpeerreviewmaketitle

\section{Introduction}
\IEEEPARstart{C}{ompared} to terrestrial wireless channels suffering from severe path loss and multi-path, the high altitude of UAVs generally leads to more dominant line-of-sight
(LoS) channels and thus largely improves the communication performance. However, the strong LoS links also make the UAV more vulnerable to attacks from terrestrial node, e.g., eavesdropping, jamming, and so on \cite{3,4}.  

On the other hand, intelligent reflecting surface (IRS)  has been proposed recently as a promising technology to improve the spectrum and energy efficiency of future wireless networks \cite{5,6}. Specifically, IRS is a planar surface which comprises a large number of reconfigurable passive reflecting elements. By adjusting the phase shifts of all reflecting elements, the reflected signals can add coherently with the signals from other paths at the intended receiver to improve the received signal power,  and destructively at the undesired receiver to suppress the interference or enhance the security \cite{7}. Therefore, IRS has been extensively studied under various wireless system setups, such as cognitive radio \cite{8,9}, simultaneous wireless information and power transfer (SWIPT) \cite{10,11}, secrecy communications \cite{12,13}, and so on. 

Thanks to its strong capacity of controlling wireless channels, IRS has great potential in tackling the security challenge in UAV communications. For example, by jointly optimizing the UAV trajectory and IRS passive beamforming, the achievable secrecy rate can be significantly improved \cite{14}. Also, it should be noted that besides eavesdropping, jamming is anther severe threat to the wireless transmission due to the openness of wireless channels. In \cite{15} and \cite{16}, it shows that by exploiting the IRS to mitigate the jamming signal from the malicious jammer, much higher throughput of the legitimate communication can be achieved. However, theses two works just focused on the terrestrial communication system in presence of jammers. When considering incorporating the UAV's flexibility  with the IRS passive beamforming to enhance the anti-jamming performance, the formulated problem becomes more complex and difficult to solve. Thus, it still remains an open problem and needs further study.

\begin{figure}
	\centering
	\includegraphics[width=6.5cm]{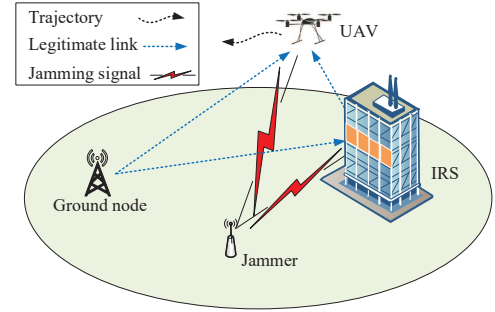}
	\caption{IRS-assisted UAV communication in the presence of a jammer.} \label{Model}
\end{figure}

Motivated by the above, in this letter we investigate the uplink transmission in an IRS-assisted UAV communication system in the presence of a malicious jammer, as shown in Fig. 1. Specifically, we aim to maximize the average rate from the ground node (GN) to UAV via the joint design of the UAV's trajectory, GN's power allocation and IRS's passive beamforming. The formulated problem is difficult to solve due to the non-convex objective function and coupled optimization variables. To tackle this challenge, we propose an alternating optimization (AO) based algorithm with the help of successive convex approximation (SCA) and semidefinate relaxing (SDR) techniques. Numerical results show that our proposed joint design algorithm significantly improves the uplink average rate compared with the benchmark algorithms. Moreover, it also shows that deploying the IRS near the jammer is more favorable to suppress the jamming signal and thus achieves better performance than deploying it near the GN when jamming power is large and the number of IRS elements is relatively small.
 
\section{System Model and Problem Formulation}
\label{sec:System Model}

In this letter, a UAV communication system is considered as shown in Fig. 1, where an IRS is deployed to assist in the transmission from a GN to a UAV in the presence of a jammer. All communication nodes are placed in the three dimensional (3D) Cartesian coordinates. The position of the jammer, GN is expressed as ${\bf{q_M}} = [{x_M},{y_M},0]$, ${\bf{q_G}} = [{x_G},{y_G},0]$. The UAV is assumed to fly at a fixed altitude ${H_0}$. The flying time of the UAV is $T$. For ease of handling, $T$ is divided into $N$ time slots, i.e., $T = N{\delta _t}$, where ${\delta _t}$ is the length of a time slot. Therefore, the trajectory of the UAV can be expressed by ${\bf{q[n]}} = {[x[n],y[n],H_0]^T},n \in {\cal{N}} = \{ 1,2,...,N\}$, ${\bf{Q}} \buildrel \Delta \over = \{ {\bf{q[n]}},\forall n\}$, which meets the mobility constraints as 
{\setlength\abovedisplayskip{1.5pt}
	\setlength\belowdisplayskip{1.5pt}
\begin{equation}\label{eq0}
{\bf{q}}\left[ 0 \right] = {{\bf{q}}^{start}}, {\bf{q}}\left[ N \right] = {{\bf{q}}^{end}},\\
\end{equation}
\begin{equation}\label{eq0}
\left\| {{\bf{q}}\left[ n \right] - {\bf{q}}\left[ {n - 1} \right]} \right\| \le {V_{\max }}\delta_t,\\
\end{equation}
where ${V_{\max }}$ denote the maximum flying speed. Assuming that $P[n]$ is the transmit power of the GN in time slot $n$,  we have the following power constraints as 
{\setlength\abovedisplayskip{1.5pt}
	\setlength\belowdisplayskip{1.5pt}
\begin{equation}\label{eq0}
	\frac{1}{N}\sum\limits_{n = 1}^N {P\left[ n \right]}  \le {P_{avg}},\\
\end{equation}
\begin{equation}\label{eq0}
	P\left[ n \right] \le {P_{peak}},\forall n,	
\end{equation}
where ${P_{avg}}$ and ${P_{peak}}$ are the average transmit power and the maximum transmit power of the GN, respectively.

We assume that the UAV, jammer, and GN are equipped with a single antenna, while the IRS is equipped with a uniform planar array (UPA) containing $K{\rm{ = }}{K_x} \times {K_z}$ reflecting elements in the $x - z$ plane. The grid of IRS is denoted by ${{\bf{q}}_R} = {[{x_R},{y_R},{z_R}]^T}$. We assume ${\bf{\Theta }} \buildrel \Delta \over = \left\{ {\Theta [n] = {\rm{diag}}\left( {{{\rm{e}}^{{\rm{j}}{\theta _{\rm{1}}}[{\rm{n}}]}},...,{{\rm{e}}^{{\rm{j}}{\theta _{\rm{K}}}[{\rm{n}}]}}} \right),\forall n} \right\}$ as the diagonal phase shift matrix of IRS, where ${\theta _i}[n] \in [0,2\pi )$, $i \in \{ 1,...,K\}$, is the phase shift of the $i$-th reflecting element in slot $n$. 

Due to the rare blockages in the air and the flexible deployment of IRS, we assume that all channels are LoS channels in the considered system. Specifically, the channel from the GN to the UAV (G-U channel) in time slot $n$ is expressed by
\begin{equation}
    {h_{GU}}\left[ n \right] = \sqrt {{L_{GU}}\left[ n \right]} {g_{GU}}\left[ n \right],
\end{equation}
where ${g_{GU}}\left[ n \right] = {e^{ - j\frac{{2\pi {d_{GU}}\left[ n \right]}}{\lambda}}}$ and ${L_{GU}}[n] = \rho d_{GU}^{ - 2}\left[ n \right]$ represent the phase response and path loss, respectively. Moreover, ${d_{GU}}\left[ n \right] = \left\| {{\bf{q}}\left[ n \right] - {{\bf{q}}_G}} \right\|$ is the distance between the GN and the UAV, $\lambda$ is the carrier wavelength. $\rho$ is the path loss at the reference distance ${D_0} = 1{\rm{m}}$. The same channel model is adopted for the channel from the jammer to the UAV, i.e., ${h_{MU}[n]}$. 

Further, the GN-IRS-UAV channel is then modeled as a concatenation of
three components, namely, the GN-IRS channel, IRS’s reflection with phase shifts, and IRS-UAV channel. Specifically, the IRS-UAV channel denoted by ${{\bf{h}}_{RU}}\left[ n \right] \in {\mathbb{C}}$, can be given by 
\begin{equation}
    {{\bf{h}}_{RU}}\left[ n \right] = \sqrt {{L_{RU}}\left[ n \right]} {{\bf{g}}_{RU}}\left[ n \right],
\end{equation}
where ${L_{RU}}[n] = \rho d_{RU}^{ - 2}\left[ n \right]$ denotes the passloss of the reflecting channels. Denoting ${d_{RU}}[n] = \left\| {{\bf{q}}[n] - {{\bf{q}}_R}} \right\|$ by the distance between the UAV and the IRS, the phase response of the IRS-UAV channel, i.e.,  ${{\bf{g}}_{RU}} \in {\mathbb{C}} {^{K }}$ is then given by
\begin{equation}
{{\bf{g}}_{RU}}\left[ n \right] = {e^{ - j\frac{{2\pi {d_{RU}}\left[ n \right]}}{\lambda }}}{m_x}\left[ n \right] \otimes {m_z}\left[ n \right],
\end{equation}
where 
\[{m_x}\left[ n \right]{\rm{ = }}{[1,{e^{ - j{\alpha _x}[n]}},...,{e^{ - j\left( {{K_x} - 1} \right){\alpha _x}[n]}}]^T},\]
\[{m_z}\left[ n \right]{\rm{ = }}{[1,{e^{ - j{\alpha _z}[n]}},...,{e^{ - j\left( {{K_z} - 1} \right){\alpha _z}[n]}}]^T},\]
 \[{\alpha _x}[n] = \frac{{2\pi d}}{\lambda }\sin {\phi _{RU}}\left[ n \right]\cos {\varphi _{RU}}\left[ n \right],\]
\[{\alpha _z}[n] = \frac{{2\pi d}}{\lambda }\sin {\phi _{RU}}\left[ n \right]\sin {\varphi _{RU}}\left[ n \right],\] $d$ is the IRS element separation, ${\phi _{RU}}\left[ n \right]$ and ${\varphi _{RU}}\left[ n \right]$ represent the vertical and horizontal angle of arrival (AoA) at the IRS, respectively, while  
$\sin {\phi _{RU}}\left[ n \right]\cos {\varphi _{RU}}\left[ n \right] = \frac{{{H_0} - {z_R}}}{{{d_{RU}}[n]}}$,
$\sin {\phi _{RU}}\left[ n \right]\sin {\varphi _{RU}}\left[ n \right] = \frac{{x\left[ n \right] - {x_R}}}{{{d_{RU}}\left[ n \right]}}$. The GN-IRS  channel, i.e., ${\bf{h}}_{GR}^H$, is modeled by a similar procedure. Thus, the cascaded GN-IRS-UAV channel, is expressed by   
\begin{equation}
    {h_{GRU}}\left[ n \right] = {\bf{h}}_{GR}^H\left[ n \right]\Theta \left[ n \right]{{\bf{h}}_{RU}}\left[ n \right].
\end{equation}
Note that the cascaded Jammer-IRS-GN channel, i.e., ${{\bf{h}}_{MRU}}$ can be modeled as the same. By denoting ${{\rm{h}}_{G}}[n] = {{h_{GU}}\left[ n \right] + {h_{GRU}}\left[ n \right]}$ and ${{\rm{h}}_{M}}[n] = {{h_{MU}}\left[ n \right] + {h_{MRU}}\left[ n \right]}$, the received signal at the UAV in time slot $n$ is given by  
\begin{equation}
    y\left[ n \right] = \sqrt {P\left[ n \right]} {h_G}\left[ n \right]{s_G} + \sqrt {{P_M}} {h_M}\left[ n \right]{s_M} + {n_0},
\end{equation}
where ${P_M}$ denotes the transmit power of the jammer, $s_G$ and $s_M$ represent the information-carrying signal and the jamming signal with unit power, respectively, while ${n_0}$ is the additive white Gaussian noise (AWGN) with zero mean and variance ${\sigma ^2}$. Finally, the achievable average rate over the flying time $T$ is given by 
\begin{equation}\label{8}
R =\frac{1}{N} \sum\limits_{n \in {\cal{N}}} {{{\log }_2}\left( {1 + \frac{{P[n]{{\left| {{{\rm{h}}_{G}}[n]} \right|}^2}}}{{{P_M}{{\left| {{{\rm{h}}_{M}}[n]} \right|}^2} + {\sigma ^2}}}} \right)}. 
\end{equation}

We aim to maximize the $R$ via a joint design of the UAV trajectory $\bf{Q}$, GN's transmit power $\bf{P}$ and IRS phase shift matrix ${\bf{\Theta}}$. Thus, the optimization problem is formulated as 
\
\begin{equation*}
	\begin{split}{\left( {{\rm{P0}}} \right)}
		:{\rm{ }}&\mathop {\max}\limits_{\bf{P},\bf{Q},\bf{\Theta} } {R}\\
		{\rm{      }}&{\rm s.t}. \quad {\theta _i}[n] \in [0,2\pi ),i \in \{ 1,...,K\} ,\forall n,\\
		&~~~~~~\left( {\rm{1}} \right),\left( {\rm{2}} \right),\left( {\rm{3}} \right),\left( {\rm{4}} \right).		
	\end{split}
\end{equation*}

It is challenging to solve ${\left( {{\rm{P0}}} \right)}$ due to the non-convex objective function and the coupled optimization variables. However, it can be effectively solved by dividing the problem into three sub-problems by applying the block coordinate descent (BCD) method. This conducts us to propose an algorithm based on alternating optimization (AO), which solves suboptimally by iterating on one of the optimizations,  while fixing the other two in each iteration until convergence is achieved. 

\section{The Proposed Alternating Algorithm}
\label{The Proposed Alternating Algorithm}

\vspace{3pt}
{\subsection{ Sub-Problem 1: Optimizing ${\bf{P}}$ for Given ${\bf{Q}}$ and ${\bf{\Theta}}$ }}

For given the UAV trajectory ${\bf{Q}}$ and IRS phase shift matrix $\bf{\Theta} $, the problem (P0) can be expressed as
\vspace{3pt}
\begin{equation*}
	\begin{split}{\left( {{\rm{P1}}} \right)}
	    :{\rm{ }}&\mathop {\max }\limits_{\bf{P}} \frac{1}{N}\sum\limits_{n \in {\cal{N}}} {{{\log }_2}\left( {1 + \frac{{P[n]{{\left| {{{\rm{h}}_{G}}[n]} \right|}^2}}}{{{P_M}{{\left| {{{\rm{h}}_{M}}[n]} \right|}^2} + {\sigma ^2}}}} \right)}\\
	    &{\rm{s}}{\rm{.t}}{\rm{.}}\left( {\rm{3}} \right), \left( {\rm{4}} \right).
   \end{split}
\end{equation*}
This is a standard convex optimization problem that can be efficiently solved by CVX.

{\subsection{Sub-Problem 2: Optimizing ${\bf{\Theta}}$ for Given ${\bf{Q}}$ and ${\bf{P}}$}}

For given trajectory ${\bf{Q}}$ and transmit power ${\bf{P}}$, by denoting
\begin{equation*}
{\mathbf{g}}_J^H\left[ n \right]{G_J}[n]v[n] = {h_{JU}}[n] + h_{JRU}[n],
\end{equation*}
 where \[{{\bf{G}}_J}[n] = {\rm{diag}}\left( {\left[ {\begin{array}{*{20}{c}}
 			{{{\bf{h}}_{RU}}[n]}&{{h_{JU}}[n]}
 	\end{array}} \right]} \right),\] \[{{\bf{g}}_J}[n] = {[{\bf{h}}_{JR}^H[n],1]^H}, J \in \left\{ {G,M} \right\},\] and ${\bf{v}}[n] = [{e^{j{\theta _1}[n]}},...,{e^{j{\theta _K}[n]}},1]$, ${\left( {{\rm{P0}}} \right)}$ can be transformed into  
\begin{equation*}\label{12}\small
\begin{split}{\left( {{\rm{P2}}} \right)}
	:{\rm{ }}&\mathop {\max }\limits_{\bf{v}}  \frac{1}{N}\sum\limits_{n \in {\cal{N}}} {{{\log }_2}\left( {1 + \frac{{P[n]{{\left| {{\mathbf{g}}_G^H\left[ n \right]{{{\bf{G}}_G}}[n]{\bf{v}}[n]} \right|}^2}}}{{{P_M}{{\left| {{\mathbf{g}}_M^H\left[ n \right]{{\bf{G}}_M}[n]{\bf{v}}[n]} \right|}^2} + {\sigma ^2}}}} \right)} \\
	&{\rm{s}}{\rm{.t}}{\rm{.}}\quad {\theta _i}[n] \in [0,2\pi ),i \in \{ 1,...,K\} ,\forall n.
\end{split}
\end{equation*}
Further, we have 
\vspace{2pt}
\begin{equation*}
	{\left| {{\bf{g}}_J^H[n]{{\bf{G}}_J}[n]{\bf{v}}[n]} \right|^2} =  {\rm{tr}}\left( {{{\bf{R}}_J}[n]\bf{V}[n]} \right),
\end{equation*}
where
\vspace{2pt}
\begin{equation*}
	{{\bf{R}}_J}[n] ={\mathbf{G}}_J^H\left[ n \right]{{\bf{g}}_J}[n]{\bf{g}}_J^H[n]{{\bf{G}}_J}[n],
\end{equation*}
\begin{equation*}
	{\bf{V}}[n] = {\bf{v}}[n]{{\bf{v}}^H}[n], J \in \left\{ {G,M} \right\}
\end{equation*}
${\bf{V}}[n]$ follows that  $\mathbf{V}\left[ n \right]\underset{\raise0.3em\hbox{$\smash{\scriptscriptstyle-}$}}{\succ }0$ and ${\text{rank}}\left( {{\mathbf{V}}\left[ n \right]} \right) = 1$. Since the rank-1 constraints are non-convex, we apply the SDR to relax these constraints. Thus, (P2.1) can be reformulated as
\vspace{3pt}
\begin{equation*}\label{13}
\begin{split}{\left( {{\rm{P2.1}}} \right)}
	:{\rm{ }}&\mathop {\max }\limits_{\bf{V}} \frac{1}{N}\sum\limits_{n \in {\cal{N}}} {{{\log }_2}\left( {1 + \frac{{P[n]{\rm{tr}}({{\bf{R}}_G}[n]{\bf{V}}[n])}}{{{P_M}{\rm{tr}}({{\bf{R}}_M}[n]{\bf{V}}[n]) + {\sigma ^2}}}} \right)} \\
	&{\rm{s}}{\rm{.t}}{\rm{.  }}\mathbf{V}\left[ n \right]\underset{\raise0.3em\hbox{$\smash{\scriptscriptstyle-}$}}{\succ }0, \forall n,\\
&~~~~{{\bf{V}}_{r,r}}[n] = 1,r = 1,...,K + 1,\forall n,\\
\end{split}
\end{equation*}
this problem is non-convex and is not easy to solve directly. 
Our goal is to achieve greater average rate in each time slot by finding a suitable set of phase-shift, as long as the SNR is maximized. To make it easier to solve, in every time slot where the value of the transmit power is not zero, (P2.2) can be equivalent to finding
\vspace{2pt}
\begin{equation*}\label{14}
\begin{split}{\left( {{\rm{P2.2}}} \right)}
	:{\rm{ }}&\mathop {\min }\limits_{\bf{V}} \frac{{{P_M}{\rm{tr}}({{\bf{R}}_M}[n]{\bf{V}}[n]) + {\sigma ^2}}}{{P[n]{\rm{tr}}({{\bf{R}}_G}[n]{\bf{V}}[n])}},\forall n\\
	&{\rm{s}}{\rm{.t}}{\rm{.}}\begin{array}{*{20}{c}}
		{{\mathbf{V}}\left[ n \right]\underset{\raise0.3em\hbox{$\smash{\scriptscriptstyle-}$}}{ \succ } 0,} \forall n,
	\end{array}\\
&~~~~~{{\bf{V}}_{r,r}}[n] = 1,r = 1,...,K + 1,\forall n,\\
\end{split}	
\end{equation*}
(P2.2) belongs to the combination of fractional programming and SDR. Then, we introduce ${\bf{\mu }} = \{ \mu [n]\ge 0,\forall n\}$ as the optimal value set of the objective function of (P2.2). Thus, the problem is transformed into
\vspace{3pt}
\begin{equation*}\label{15}
\begin{gathered}
	{\left( {{\rm{P2.3}}} \right)}\mathop {\min }\limits_{\mathbf{V}} 
		{{P_M}{\text{tr}}({{\mathbf{R}}_M}[n]{\mathbf{V}}[n]) + \!{\sigma ^2}}\! \!-\! \mu [n]P[n]{\text{tr}}({{\mathbf{R}}_G}[n]{\mathbf{V}}[n]) \hfill \\
	~~~~~~~~{\text{s}}.{\text{t}}.{{\mathbf{V}}_{r,r}}[n] = 1,r = 1,...,K + 1,\forall n. \hfill \\ 
\end{gathered} 
\end{equation*}
Denoting  the optimal value of (P2.3) by $\varphi \left( \mu  \right)$, ${\mathbf{V}}$ can be solved as follows. First, initialize ${\mathbf{V}}$ as ${\mathbf{\tilde V}}$, then we can obtain $ \mu[n]\!=\!\tilde \mu  $ by solving $ \varphi \left( \mu  \right) \!=\! 0 $. Next, for given $\mu [n]\!=\!\tilde \mu $, (P2.3) is an SDR problem, which thus can be efficiently solved by using CVX. Finally,  ${\mathbf{V}}$ can be obtained by repeating the above two steps until convergence. To summarize, the iterative algorithm to solve (P2.1) is given in Algorithm 1. According to \cite{13}, the objective value of (P2.1) is non-decreasing after each iteration of Algorithm 1 and has a finite upper bound. Therefore, Algorithm 1 always converges.

It should be noted that the optimal target value of (P2.1) only serves the upper bound of the (P2) since SDR is applied, and thus there is no guarantee that the obtained ${\mathbf{V}[n]}$ in each time slot is of rank-1. Specifically, if the obtained ${\mathbf{V}[n]}$ is of rank-1, it can be written as ${\bf{V}}[n] = {\bf{v}}[n]{{\bf{v}}^H}[n]$ by applying eigenvalue decomposition, and the obtained ${\bf{v}}[n]$ is the optimal solution to (P2.1). Otherwise, Gaussian randomization is needed for recovering ${\bf{v}}[n]$ approximately \cite{7}.
\begin{algorithm}[t]
	\renewcommand{\algorithmicrequire}{\textbf{Input:}}
	\renewcommand{\algorithmicensure}{\textbf{Output:}}
	\caption{An alternating algorithm for solving (P2.1)}\label{Algorithm1}
	\begin{algorithmic} [1]
		\STATE\textbf{Initialization:}
		Initialize ${\mathbf{V}}$ as ${\mathbf{\tilde V}}$.
		\STATE\textbf{repeat}
		\STATE\quad For given $ {\mathbf{V}} \!=\! {\mathbf{\tilde V}} $, obtain $ \mu[n]\!=\!\tilde \mu  $ by solving $ \varphi \left( \mu  \right) \!=\! 0 $.
		\STATE\quad For given $ \mu[n]\!=\!\tilde \mu  $, obtain $ {\mathbf{V}} \!=\! {\mathbf{\tilde V}} $ by solving (P2.3).
		\STATE{\textbf{until} the fractional increase of the objective value is below a small threshold ${\varepsilon_1 }$.  }	
	\end{algorithmic}
\end{algorithm}

{\subsection{Sub-Problem 3: Optimizing ${\bf{Q}}$ for Given ${\bf{P}}$ and ${\bf{\Theta}}$}}
For given transmit power ${\bf{P}}$ and IRS phase shift matrix $\bf{\Theta} $, we can express (P0) as
\begin{equation*}
	\begin{split}{\left( {{\rm{P3}}} \right)}
		:{\rm{ }}&\mathop {\max }\limits_{\mathbf{P}} \frac{1}{N}\sum\limits_{n \in N} {{{\log }_2}\left( {1 + \frac{{P[n]{{\left| {{{\text{h}}_G}[n]} \right|}^2}}}{{{P_M}{{\left| {{{\text{h}}_M}[n]} \right|}^2} + {\sigma ^2}}}} \right)} \\
		{\rm{      }}&{\rm s.t}. \left( {\rm{1}} \right),\left( {\rm{2}} \right).		
	\end{split}
\end{equation*}
(P3) is challenging to solve due to the non-convex objective function. It is observed that ${\mathbf{g}}_{GU}\left[ n \right]$, ${\mathbf{g}}_{MU}\left[ n \right]$, ${\mathbf{g}}_{RU}\left[ n \right]$ are complex and non-linear
with respect to the UAV trajectory variables, which makes the UAV trajectory design intractable. To overcome the difficulty, we use the UAV trajectory of the ($i-1$)th iteration to obtain an approximate ${{{\bf g}}_{GU}^{\left( {i } \right)}\left[ n \right]}$, ${{{\bf g}}_{MU}^{\left( {i } \right)}\left[ n \right]}$, ${{{\bf g}}_{RU}^{\left( {i } \right)}\left[ n \right]}$ in the $i$th iteration \cite{14}. Thus, by denoting
\vspace{2pt}
\begin{equation*}
	{{\bf{h}}_{QJ}[n]} = {[\sqrt \rho  {\rm{g}}_{JU}^{\left( i-1\right)}[n],\rho  {d_{JR}^{ - 1}\left[ n \right]} {\bf{g}}_{JR}^H[n]\Theta [n]{{\bf g}_{RU}^{\left( {i - 1} \right)}\left[ n \right]}]},
\end{equation*}
\begin{equation*}
	{{\bf{r}}_{J}}\left[ n \right] = {\left[ { {d_{JU}^{ - 1}\left[ n \right]} , {d_{RU}^{ - 1}\left[ n \right]} } \right]^T}, J \in \left\{ {G,M} \right\},
\end{equation*}
the objective function can be rewritten as
\vspace{2pt}
\begin{equation}\small\label{sub3}
	\begin{split}
	&\mathop {\max}\limits_{\bf{Q} } { \frac{1}{N}\sum\limits_{n \in {\cal{N}}} {{{\log }_2}\left( {1 + \frac{{P[n]{\bf{r}}_{G}^T[n]{\bf{h}}_{QG}^H[n]{{\bf{h}}_{QG}}[n]{{\bf{r}}_{G}}[n]}}{{{P_M}{\bf{r}}_{M}^T[n]{\bf{h}}_{QM}^H[n]{{\bf{h}}_{QM}[n]}{{\bf{r}}_{M}}[n] + {\sigma ^2}}}} \right)}}\\	
	\end{split}
\end{equation}
However, (\ref{sub3}) is still non-convex.
By introducing the relaxation variable ${\bf{L}} = \{ L[n],\forall n\} ,{\bf{I}} = \{ I[n],\forall n\} $, the original problem (P3) can be rewritten as
\vspace{3pt}
\begin{equation*}\label{p17}
\begin{split}{\left( {{\rm{P3.1}}} \right)}
	:{\rm{}}&\mathop {\max }\limits_{\bf{Q},L,I,\eta } \eta \\
	&{\rm{s}}{\rm{.t}}{\rm{.}}\frac{1}{N}\sum\limits_{n \in {\cal{N}}} {{{\log }_2}\left( {1 + \frac{1}{{L[n]I[n]}}} \right)}  \ge \eta, \\
~~~~
&~~~~P[n]{\bf{r}}_{G}^T[n]{\bf{h}}_{QG}^H[n]{{\bf{h}}_{QG}}[n]{{\bf{r}}_{G}}[n] \ge {L^{ - 1}}[n],\forall n,\\
~~~~	&~~~~{P_M}{\bf{r}}_{M}^T[n]{\bf{h}}_{QM}^H[n]{{\bf{h}}_{QM}}[n]{{\bf{r}}_{M}}[n] + {\sigma ^2} \le I[n],\forall n,\\
~~~	&~~~~\left( 1 \right),\left( 2 \right).
\end{split}		
\end{equation*}

(P3) and (P3.1) share the same optimal solution when the constraints hold with equalities \cite{3}. By applying SCA, the first constraint in (P3.1) is rewritten as
\vspace{3pt}
\begin{equation*}\label{18}
\begin{split}
	{\rm{}}&{\tilde{R}}\left( {L[n],I[n]} \right) = {\rm{lo{g_2}}}(1 + \frac{1}{{{L_0}[n]{I_0}[n]}})\\
&~~~~~~~~~~~~~~~~	+ A[n]\left( {L[n] - {L_0}[n]} \right) + B[n]\left( {I[n] - {I_0}[n]} \right),
\end{split}
\end{equation*}
where 
\vspace{3pt}
\begin{equation*}
	A[n] =  - {\log _2}\left( {\frac{e}{{{L_0}[n] + L_0^2\left[ n \right]{I_0}[n]}}} \right),
\end{equation*}
\begin{equation*}
	B[n] =  - {\log _2}\left( {\frac{e}{{{I_0}[n] + I_0^2\left[ n \right]{L_0}[n]}}} \right),
\end{equation*}
while ${L_0}\left[ n \right]$ and ${I_0}\left[ n \right]$ denote the feasible points of the first-order Taylor expansion. Further, to handle the second and the third non-convex constraints in  (P3.1),  we denote ${\bf{u}} = \{ {u}[n],\forall n\} $, ${\bf{e}} = \{ {e}[n],\forall n\} $, ${\bf{s}} = \{ {s}[n],\forall n\} $, ${\bf{t}} = \{ t[n],\forall n\} $, ${\tilde {\bf r}_{G}} = {[u[n],e[n]]^T}$, ${\tilde {\bf r}_{M}} = {[s[n],t[n]]^T}$. As such, (P3.1) can be transformed into
\vspace{5pt}
\begin{equation*}\label{19}
\begin{split}{\left( {{\rm{P3.2}}} \right)}
	:{\rm{}}&\mathop {\max }\limits_{\bf{Q},L,I,\eta \hfill\atop
	 ~	u,e,s,t\hfill} \eta \\
	{\rm{s.t.}}{\rm{}}&
	\frac{1}{N}\sum\limits_{n \in {\cal{N}}} {{\tilde{R}}(L[n],I[n]) \ge \eta ,}\\
	&P[n]\tilde {\bf{r}}_{G}^T[n]{\bf{h}}_{QG}^H[n]{{\bf{h}}_{QG}}[n]{{\tilde {\bf{r}}}_{G}}[n] \ge {L^{ - 1}}[n],\forall n,\\
	&{P_M}\tilde {\bf{r}}_{M}^T[n]{\bf{h}}_{QM}^H[n]{{\bf{h}}_{QM}}[n]{{\tilde {\bf{r}}}_{M}}[n] + {\sigma ^2} \le I[n],\forall n,\\
	& e[n] \le d_{RU}^{ - 1}\left[ n \right] \le s[n], \forall n,
	\\
	&d_{GU}^{ - 1}\left[ n \right]  \ge u[n],
	 d_{MU}^{ - 1}\left[ n \right] \le t[n], \forall n,\\
	&(1),(2).\\ 
\end{split}
\end{equation*}

\vspace{1pt}
The constraints associated with the distances $d_{GU}$, $d_{RU}$ and $d_{MU}$ are non-convex, and we rewrite them as follows
\vspace{2pt}
\begin{equation*}
	\begin{split}
&{\underbrace {{x^2}[n] + x_G^2 + {y^2}[n] + y_G^2 - 2{x_G}x[n] - 2{y_G}y[n] + {H_0}^2}_{{F_1}}}\\
& - {u^{ - 2}[n]} \le 0,\forall n,\\
\end{split}
\end{equation*}
\begin{equation*}
	\begin{split}
		&\underbrace {{{\left( {x\left[ n \right] - {x_R}} \right)}^2} \!+\! {{\left( {y\left[ n \right] - {y_R}} \right)}^2} \!+\! {{\left( {H_0 - {z_R}} \right)}^2}}_{{F_2}} \!-\! {e^{ - 2}[n]} \le 0,\forall n,\\
		&\underbrace {{{s^{ - 2}[n]}} - x_R^2 - y_R^2 - z_R^2 + 2{x_R}x\left[ n \right] + 2{y_R}y\left[ n \right] + 2{z_R}H_0}_{{F_3}}\\
		& - {x^2}[n] - {y^2}[n] - {{H_0}^2}\le 0,\forall n,\\
		&\underbrace {{{t^{ - 2}[n]}} \!-\! x_M^2 - y_M^2 + 2{x_M}x[n] + 2{y_M}y[n]}_{{F_4}} - {x^2}[n]- {y^2}[n]\\
		& - {{H_0}^2}\le 0,\forall n.\\
	\end{split}
\end{equation*}
It is obvious that a part of the above constraints comprising non-convex terms. And we use the first-order Taylor expansion of non-convex terms to deal with these cases. Specifically, by denoting ${{\bf{x}}_0} = \{ {x_0}[n]\} _1^N,{{\bf{y}}_0}{\rm{ = }}\{ {y_0}[n]\} _1^N,{{\bf{u}}_0} = \{ {u_0}[n]\} _1^N,{{\bf{e}}_0} = \{ {r_0}[n]\} _1^N,{{\bf{s}}_0} = \{ {s_0}[n]\} _1^N,{{\bf{t}}_0} = \{ {t_0}[n]\} _1^N$ and ${{\bf{r}}_{G,0}} = \{ {\tilde r_{G,0}}[n]\} _{n = 1}^N$, the above distance constraints can be transformed into convex, written as
\vspace{5pt}
\begin{equation*}
	{C_1}:\left\{ \begin{gathered}
		{F_1} - {{u_0^{ - 2}}[n]} + 2{{u_0^{ - 3}}[n]}(u[n] - {u_0}[n]) \leqslant 0, \hfill \\
		{F_2} - {{e_0^{ - 2}}[n]} + 2{{e_0^{ - 3}}[n]}(e[n] - {r_0}[n]) \leqslant 0, \hfill \\
		{F_3} + x_0^2\left[ n \right] - 2{x_0}\left[ n \right]x\left[ n \right] + y_0^2\left[ n \right] - 2{y_0}\left[ n \right]y\left[ n \right] \leqslant 0, \hfill \\
		{F_4} + x_0^2\left[ n \right] - 2{x_0}\left[ n \right]x\left[ n \right] + y_0^2\left[ n \right] - 2{y_0}\left[ n \right]y\left[ n \right] \leqslant 0. \hfill \\ 
	\end{gathered}  \right.
\end{equation*}
Therefore, (P3.2) can be rewritten by
\begin{equation*}\small\label{20}
\begin{split}{\left( {{\rm{P3.3}}} \right)}
	:{\rm{}}&\mathop {\max }\limits_{\bf{Q},L,I,\eta \hfill\atop
		u,e,s,t\hfill} \eta \\
	{\rm{s}}{\rm{.t}}{\rm{.}}&P\left[ n \right](2\Re \left[ {\widetilde {\text{r}}_{G,0}^T[n]{\mathbf{h}}_{QG}^H[n]{{\mathbf{h}}_{QG}}[n]{{\widetilde {\text{r}}}_G}[n]} \right] \hfill \\
	&~~~~~~- {\widetilde {\text{r}}_{G,0}}^T[n]{\mathbf{h}}_{QG}^H[n]{{\mathbf{h}}_{QG}}[n]{\widetilde {\text{r}}_{G,0}}[n]) \geqslant {L^{ - 1}}[n],\forall n, \hfill \\ 
&{{P_M}\tilde {\rm{r}}_{M}^T[n]{\bf{h}}_{QM}^H[n]{{\bf{h}}_{QM}}[n]{{\tilde {\rm{r}}}_{M}}[n] + {\sigma ^2} \le }I[n],\forall n,\\
      &\frac{1}{N}\sum\limits_{n \in {\cal{N}}} {{\tilde{R}}(L[n],I[n]) \ge \eta ,\forall n,}\\
      &C_1,(1),(2).
\end{split}	
\end{equation*}
which is a convex optimization problem, and thus can be solved with the CVX.
\begin{algorithm}[t]
	\renewcommand{\algorithmicrequire}{\textbf{Input:}}
	\renewcommand{\algorithmicensure}{\textbf{Output:}}
	\caption{An alternating algorithm for solving {(\rm{P0}})}\label{Algorithm1}
	\begin{algorithmic} [1]
		\STATE\textbf{Initialization:}
		~Set~the iteration number $i = 0,$ and an initial solution $\left( {{\bf{\Theta} }^{(i)},{\bf{P}}^{(i)},{\bf{Q}}^{(i)}} \right).$
		\STATE\textbf{repeat.}
		\STATE\quad ~Update ${{\bf{P}}^{\left( i \right)}}$ to ${{\bf{P}}^{\left( {i{\rm{ + }}1} \right)}}$ by solving {(\rm{P1})} with given ${{\bf{Q}}^{\left( i \right)}}$ and ${\Theta ^{(i)}}$.
		\STATE\quad ~Update ${{\bf{\Theta}}^{\left( i \right)}}$ to ${{\bf{\Theta}}^{\left( {i{\rm{ + }}1} \right)}}$ by solving {(\rm{P2.1})} with given ${{\bf{Q}}^{\left( {i{\rm{ + }}1} \right)}}$ and ${{\bf{P}}^{\left( i \right)}}$.	
		\STATE\quad ~Update ${{\bf{Q}}^{\left( i \right)}}$ to ${{\bf{Q}}^{\left( {i{\rm{ + }}1} \right)}}$ by solving {(\rm{P3.3})} with given ${{\bf{P}}^{\left( {i{\rm{ + }}1} \right)}}$ and ${{\bf{\Theta}}^{\left( i \right)}}$.
		\STATE{\quad Update $ i \leftarrow i + 1\ $}.
		\STATE{\textbf{until}~ the fractional increase of the objective value is below a small threshold $\varepsilon_2$.  }	
	\end{algorithmic}
\end{algorithm}
{\subsection{ Overall Algorithm }}
The proposed overall algorithm is summarized in Algorithm 2. The main complexity of the overall algorithm lies in solving (P2.1) and (P3.3). Specifically, the complexity of solving (P2.1) by applying Algorithm 1 is given by $
{\cal O}\left(I_1  {\sqrt {K + 1} \left( {N{{\left( {K + 1} \right)}^3} + {N^2}\left( {K + 1} \right) + {N^3}} \right)}  \right) $, where $I_1$ is the number of iterations. On the other hand, the complexity of solving (P3.3) is given by ${\cal O}\left( {{\left( {9N} \right)}^{3.5}}\right)$. Therefore, the overall computational complexity is  ${\cal O}\!\left(\! {{I_1}{I_2}\sqrt {K \!+\! 1} \!\left(\! {N{{\left( {K \!+\! 1} \right)}^3} \!+\! {N^2}\left( {K \!+\! 1} \right) \!+\! {N^3}} \right) \!+\! {I_2}{{\left( {9N} \right)}^{3.5}}} \right)$, where $I_2$ is the number of iterations required for solving (P0).

\begin{figure*}[t]
	\centering
	\subfigure[UAV’s trajectory]{
		\begin{minipage}{0.3\linewidth}
			\centering
			\includegraphics[width=2.5in,height=2.35in]{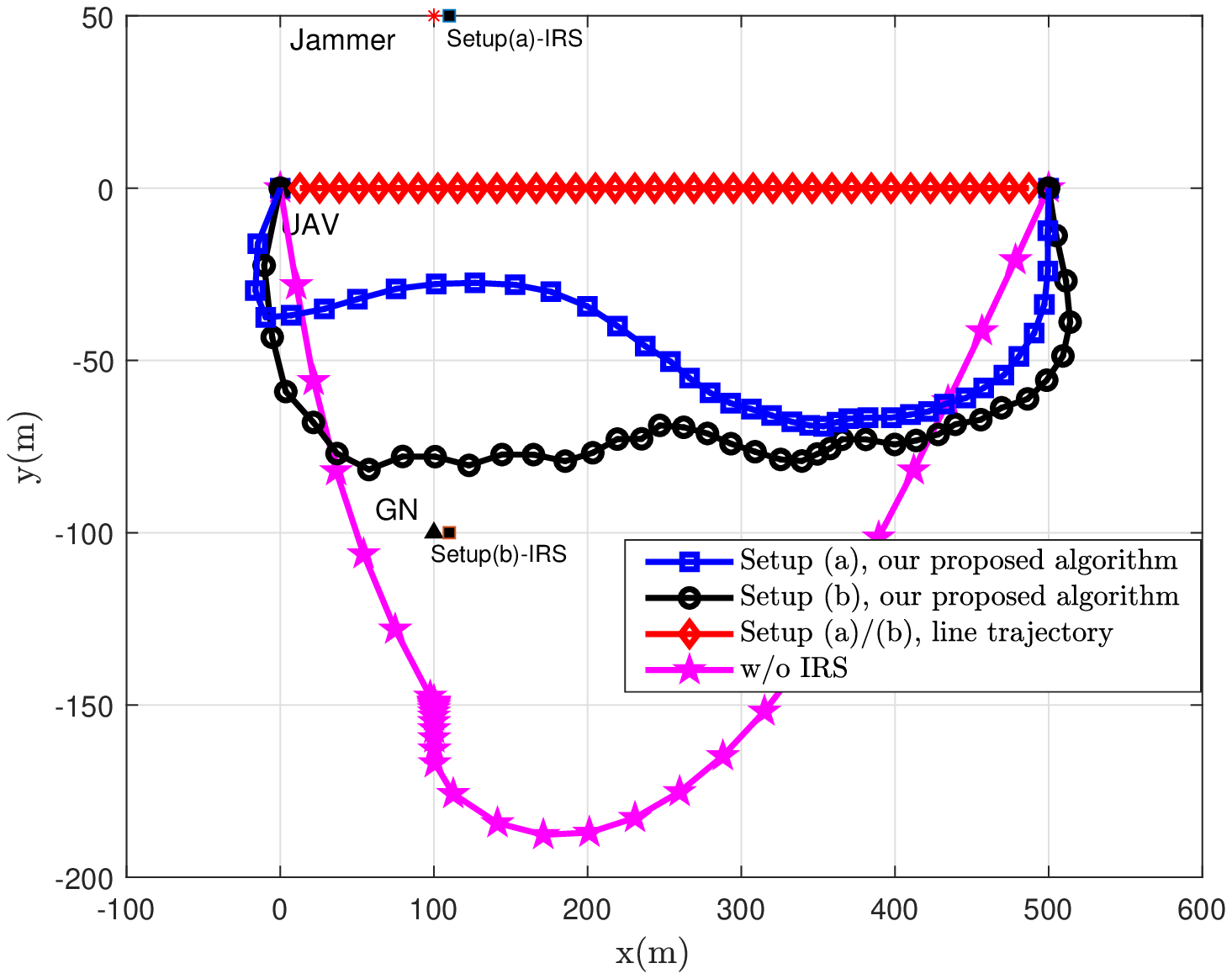}
		\end{minipage}
	} \hspace{3mm}
	\subfigure[Average rate versus $P_M$]{
		\begin{minipage}{0.3\linewidth}
			\centering
			\includegraphics[width=2.5in,height=2.35in]{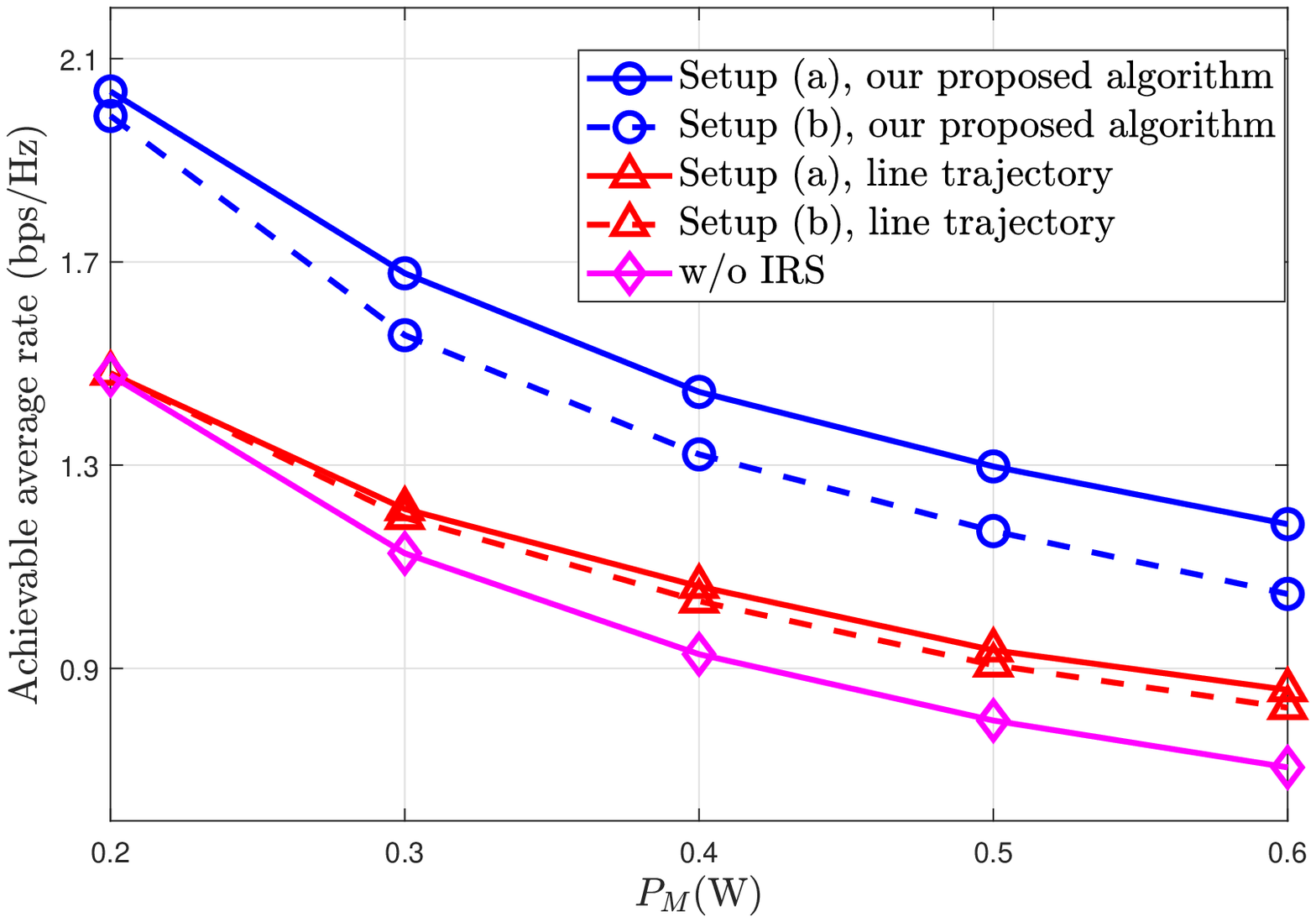}
		\end{minipage}
	}  \hspace{3mm}
	\subfigure[Average rate versus $K$]{
		\begin{minipage}{0.3\linewidth}
			\centering
			\includegraphics[width=2.5in,height=2.35in]{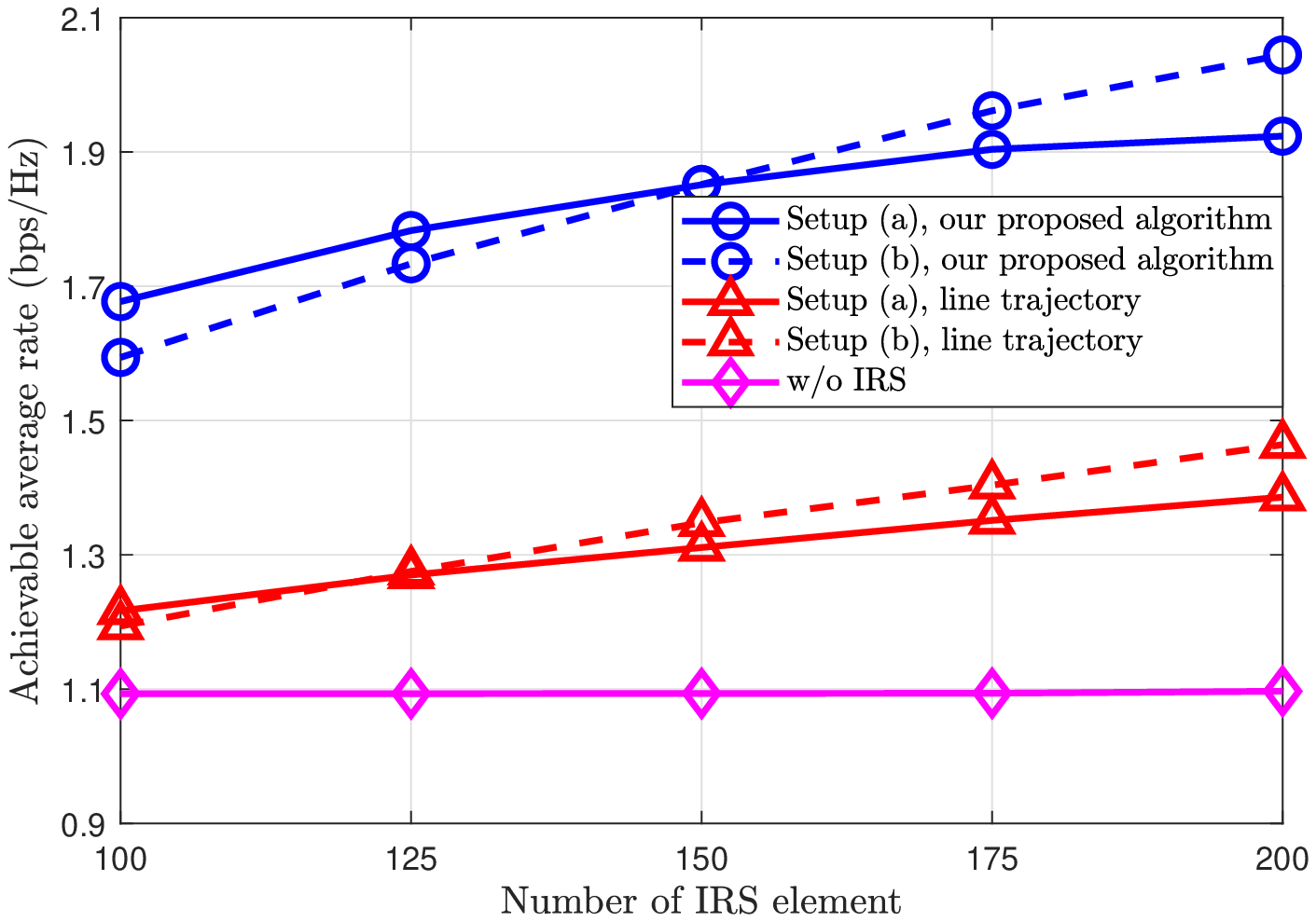}
		\end{minipage}
	} 
	\caption{Comparison between the proposed and benchmark schemes.}
	\label{fig2}
\end{figure*}

\vspace{10pt}
\section{Numerical Results}
\label{Numerical Results}
\vspace{2pt}
To study the impacts of the deployment of IRS, we consider two different setups. In particular, for Setup (a), the IRS is deployed at (110, 50, 5), i.e., nearby the jammer; while for Setup (b), the IRS is deployed at (110, -100, 5), i.e., nearby the ground node. Besides the proposed Algorithm 2, the cases with fixed line trajectory ({\bf{``line trajectory"}}) and without IRS ({\bf{``w/o IRS"}}) are also considered for performance comparison. The parameters are set as the same as in \cite{3}: ${{\rm{{\bf{q}}}}^{start}} = \left( {0,0,100} \right)$, ${{\rm{{\bf{q}}}}^{end}} = \left( {500,0,100} \right)$, ${{\rm{{\bf{q}}}}_M} = \left( {100,50,0} \right)$, ${{\rm{{\bf{q}}}}_G} = \left( {100,-100,0} \right)$, $H_0 = 100$ m, ${V_{\max }} = 60$, ${P_M} = 0.4$ W, ${P_{avg}} = 0.2$ W, ${P_{peak}} = 0.5$ W, $\rho  = {10^{ - 3}}$, ${\sigma ^2} =  - 140$ dbm/Hz, $K = 50$, ${\delta _t} = 0.5$, $\varepsilon_1=\varepsilon_2 {\rm{ = }}{10^{ - 3}}$.

Fig.\ref{fig2}(a) shows the UAV's trajectory in different cases. Our ultimate goal is to increase the average rate of the communication system. In the absence of IRS, the UAV approaches the GN along the line trajectory to enhance the transmission of information while also avoiding the jammer as possible as it can. It can be observed that the trajectory in our proposed algorithm can significantly decrease the flying path length of the UAV compared to the case without IRS. This is because the proposed algorithm balances the channel gains between the direct channels and reflecting channels in each time slot in order to choose a trajectory, so as to achieve the best average rate. In addition, we can observe that in Setup (a), the IRS can greatly reduce the jamming, thus the UAV can be closer to the line trajectory compared with Setup (b).

Fig.\ref{fig2}(b) plots the average rate of proposed algorithm in two setups versus jamming power under $K=100$. It is observed that by deploying the IRS, the average rate can be increased, even with a fixed line trajectory. The reason is that IRS can enhance the information-carrying signals and reduce the jamming signal by passive beamforming in Setup (a) and Setup (b), respectively. It is also observed as compared with the ``line trajctory" algorithm, our proposed algorithm achieves much higher average rates in both setups due to the joint passive beamforming design with trajectory optimization.  Moreover, one can observe that as the jamming power increases, the performance gap between two setups becomes larger in both our proposed algorithm and the ``line trajectory" algorithm. This is because the received signal at the UAV tends to be interference-dominant for high jamming power, and thus deploying the IRS nearby the jammer for interference reduction is more effective than deploying it nearby the GN for signal enhancement.

Fig.\ref{fig2}(c) plots the average rate of proposed algorithm in two setups versus the number of IRS elements $K$ under $P_M=0.3$ W. It is observed that with the increasing of $K$, the average rate for the cases with IRS all improved, which verifies the performance gain by enlarging the IRS size. It is also observed that the achievable average rate for Setup (a) is higher than that for Setup (b) first, and then becomes lower than the latter as $K$ increases. This is because when the number of IRS element is sufficiently large, the jamming signal in Setup (a) is well reduced and thus the reception at the UAV is no more interference-dominant. As a result, the performance gain from increasing $K$ for interference reduction becomes smaller. On contrast, for Setup (b), the reception at the UAV can substantially benefit from increasing $K$ because the IRS in this case mainly focuses on enhancing the information signal from GN.

\vspace{-5pt}
\section{Conclusions}\label{Conclusions}

In this letter, we studied the uplink UAV communication system  assisted by IRS in the presence of jammer. By considering the transmit power, IRS passive beamforming, and UAV trajectory, an alternating optimization algorithm was proposed to solve the rate maximization problem by exploiting the BCD, SDA and SDR techniques. Simulation results showed that the proposed algorithm significantly improved the uplink average rate compared with the benchmark algorithms. It also showed that deploying the IRS near the jammer achieved better performance than deploying it near the GN under severe jamming with a relatively small number of IRS elements.

\bibliographystyle{IEEEtran}

\begin{thebibliography}{1}

\bibitem{4}
Y. Zeng, Q. Wu and R. Zhang, ``Accessing From the Sky: A Tutorial on UAV Communications for 5G and Beyond," \emph{Proceedings of the IEEE}, vol. 107, no. 12, pp. 2327$-$2375, Dec. 2019.
\bibitem{3}
Y. Wu, W. Fan, W. Yang, X. Sun and X. Guan, ``Robust Trajectory and Communication Design for Multi-UAV Enabled Wireless Networks in the Presence of Jammers," \emph{IEEE Access}, vol. 8, pp. 2893$-$2905, 2020.
\bibitem{5}
Q. Wu, S. Zhang, B. Zheng, C. You, and R. Zhang, ``Intelligent Reflecting Surface-Aided Wireless Communications: A Tutorial," \emph{IEEE Trans. Commun.}, vol. 69, no. 5, pp. 3313$-$3351, May. 2021.
\bibitem{6}
Q. Wu and R. Zhang, ``Towards Smart and Reconfigurable Environment: Intelligent Reflecting Surface Aided Wireless Network," \emph{IEEE Commun. Mag.}, vol. 58, no. 1, pp. 106$-$112, January 2020.
\bibitem{7}
Q. Wu and R. Zhang, ``Intelligent reflecting surface enhanced wireless network via joint active and passive beamforming,” \emph{IEEE Trans. Wireless Commun.}, vol. 18, no. 11, pp. 5394$-$5409, Nov. 2019.
\bibitem{8}
X. Guan, Q. Wu, and R. Zhang, ``Joint power control and passive beamforming in IRS-assisted spectrum sharing,”\emph{IEEE Commun. Lett.}, vol. 24, no. 7, pp. 1553$-$1557, Jul. 2020.
\bibitem{9}
D. Xu, X. Yu, and R. Schober, ``Resource allocation for intelligent reflecting surface-assisted cognitive radio networks,” \emph{Proc. IEEE SPAWC.}, May 2020.
\bibitem{10}
Q. Wu and R. Zhang, ``Joint active and passive beamforming optimization for intelligent reflecting surface assisted SWIPT under QoS constraints,” \emph{IEEE J. Sel. Areas Commun.}, vol. 38, no. 8, pp. 1735$-$1748, Aug. 2020.
\bibitem{11}
C. Pan et al., ``Intelligent reflecting surface enhanced MIMO broadcast- ing for simultaneous wireless information and power transfer,” \emph{IEEE J. Sel. Areas Commun.}, vol. 38, no. 8, pp. 1719$-$1734, Aug. 2020.
\bibitem{12}
X. Guan, Q. Wu, and R. Zhang, ``Intelligent reflecting surface assisted secrecy communication: Is artificial noise helpful or not?” \emph{IEEE Wireless Commun. Lett.}, vol. 9, no. 6, pp. 778$–$782, June 2020.
\bibitem{13}
H. Shen, W. Xu, S. Gong, Z. He, and C. Zhao, ``Secrecy rate maximization for intelligent reflecting surface assisted multi-antenna communications,” \emph{IEEE Commun. Lett.}, vol. 23, no. 9, pp. 1488$-$1492, Sep. 2019.
\bibitem{14}
S. Li, B. Duo, M. Di Renzo, M. Tao and X. Yuan, ``Robust Secure
UAV Communications with the Aid of Reconfigurable Intelligent Surfaces,”
\emph{IEEE Trans. Wireless Commun.}, vol. 20, no. 10, pp. 6402$-$6417, Oct. 2021.
\bibitem{15}
H. Yang et al., ``Intelligent Reflecting Surface Assisted Anti-Jamming
Communications: A Fast Reinforcement Learning Approach,” \emph{IEEE Trans. Wireless Commun.}, vol. 20, no. 3, pp. 1963$-$1974, March. 2021.
\bibitem{16}
Y. Sun, K. An, J. Luo, Y. Zhu, G. Zheng and S. Chatzinotas, “Intelligent
Reflecting Surface Enhanced Secure Transmission Against Both
Jamming and Eavesdropping Attacks,” \emph{IEEE Trans. Veh. Technol.}, vol. 70, no. 10, pp. 11017$-$11022, Oct. 2021.
\end{thebibliography}
\balance

\end{document}